\documentclass[aps,prb,twocolumn]{revtex4}
\usepackage{amsmath,amssymb,amsfonts,mathrsfs}
\usepackage[USenglish]{babel}
\usepackage[dvips]{graphicx}
\usepackage{subfigure}
\usepackage[usenames,dvips]{color}
\usepackage[colorlinks=true,linkcolor=blue,citecolor=red, bookmarksnumbered=true,bookmarks=true]{hyperref}
\usepackage{epsfig}
\def\eg{\mbox{\it e.g.\ }}
\def\ie{\mbox{\it i.e.\ }}
\def\=={\equiv}

\def\qed{\raise1pt\hbox{\vrule height5pt width5pt depth0pt}}

\def\cG0{{\cal G}_0}

\def\uc2{$U_{c2}$}
\def\uc1{$U_{c1}$}

\def\+{\dagger}

\begin{document}

\title{Extended Hubbard model: Charge Ordering and Wigner-Mott transition}
\author{A. Amaricci}
\affiliation{Department of Physics and Astronomy, Rutgers University, Piscataway,
New Jersey 08854, USA}

\author{A. Camjayi}
\affiliation{Department of Physics and Astronomy, Rutgers University, Piscataway,
New Jersey 08854, USA}

\author{D. Tanaskovi\'{c}}
\affiliation{Scientific Computing Laboratory, Institute of Physics Belgrade, Pregrevica
118, 11080 Belgrade, Serbia}

\author{K. Haule}
\affiliation{Department of Physics and Astronomy, Rutgers University, Piscataway,
New Jersey 08854, USA}

\author{V. Dobrosavljevi\'{c}}
\affiliation{Department of Physics and National High Magnetic Field Laboratory,
Florida State University, Tallahassee, Florida 32306, USA.}

\author{G. Kotliar}
\affiliation{Department of Physics and Astronomy, Rutgers University, Piscataway,
New Jersey 08854, USA}

\begin{abstract}
Strong correlations effects, which are often associated to the approach
to a Mott insulating state, in some cases may be observed even far
from half-filling. This typically happens whenever the inter-site
Coulomb repulsion induces a tendency towards charge ordering, an effect
that confines the electrons, and in turn favors local moment formation,
\ie Mott localization. A distinct intermediate regime then emerges
as a precursor of such a Wigner-Mott transition, which is characterized
by both charge and spin correlations, displaying large mass enhancements
and strong renormalizations of other Fermi liquid parameters. Here
we present a careful study of a quarter filled extended Hubbard model
- a simple example where such physics can be studied in detail, and
discuss its relevance for the understanding of the phenomenology of
low-density two dimensional electron gases. 
\end{abstract}


\maketitle

\section{Introduction}

\label{Sec1} 
Early theoretical and experimental investigations of
two dimensional electron gases (2DEG) largely focused on disorder
effects that dominate the so-called ``diffusive'' regime $k_{B}T\ll\hbar/\tau$,
$\tau$ being the impurity scattering rate. In this situation, which
is best realized in relatively low-mobility materials, the interaction
effects are expected to be weak and coherent multiple-scattering processes
dominate -- ultimately leading to the formation of bound electronic
states through Anderson localization. This impurity effect was first
predicted in the famed scaling theory of localization,\cite{gang4}
for non-interacting and weakly disordered 2DEG systems, and was quickly
extended to include the weak interaction corrections.~\cite{Altshuler80}
The predicted logarithmic rise of the resistivity at low temperature
was soon confirmed by experiments on thin metallic films and two-dimensional
semiconducting surfaces.~\cite{Dolan79,Bishop80} All this intensive
activity around the 2DEG contributed to the emergence of a widely
held opinion that, in these systems, even a minute amount of (impurity)
disorder will localize all the electronic states at $T=0$. If this
were true, then there should not exist a sharp metal-insualtor transition
(MIT) in any two dimensional system; the density and/or temperature
dependence of transport should simply reveal a gradual crossover from
weak to strong localization.

Given this conventional lore focusing on disorder effects, the 1994
pioneering experiment of Kravchenko et al.\cite{kravchenko1994} provided
quite a surprise. It reported low temperature transport behavior in
ultra clean 2D Silicon samples which, around this time, became available
due to technological advances in fabricating semiconductor devices.
Kravchenko's work presented first evidence for dramatic changes in
the temperature dependence of the resistivity in a narrow density
range, suggesting the possibility of a metal-insulator transition
(MIT) in the 2DEG. This result - quite surprisingly - passed almost
unnoticed for several years, until confirmed by independent transport
and magnetic response measurements on clean Si-MOSFET\cite{Popovic97},
as well as on other semiconductor heterostructures (e.g. GaAs/AlGaAs)
\cite{Hanein98}. These results have attracted a great deal of attention
to this field, triggering a surprisig variety of proposed theoretical
scenarios\cite{abrahams-rmp01}, many of which were quickly ruled
out on experimental grounds\cite{Kravchenko_Sarachik_RepProgPhys2004}.

\subsection{Strong correlations revealed}
The next ten years produced significant new information from complementary
studies by several groups, which reported singular enhancement of
the effective mass,\cite{Kravchenko_Sarachik_RepProgPhys2004} $m^{*}\sim(n-n_{c})^{-1}$,
and emphasized the central role played by spin physics in the low-density
2DEG systems. In particular, experimental measurements of the electronic
spin susceptibility in the proximity of the MIT shown a Curie-Weiss
behavior $\chi/n\simeq g\mu_{B}^{2}/T$, suggesting almost total conversion
of the electrons into local magnetic moments below the critical density.
\cite{Shashkin02,Zala01,Okamoto99} While the inital attention concentrated
on the role of disorder\cite{Punnoose}, more recent experiments\cite{Kravchenko_Sarachik_RepProgPhys2004}
made it increasingly clear that many key experimental features seem
to persist even in the cleanest samples, thus pointing to effects
intrinsic to the low-density 2DEG. All these experimental results
have contributed to validate the emerging idea~\cite{pankov-2008-77,Alberto_NaturePhysics}
that\emph{ strong interactions are the primary driving force} behind
the instability of the Fermi liquid phase in favor of an insulating
state in the low density 2DEG.

In very recent work,~\cite{pankov-2008-77,Alberto_NaturePhysics}
radically new ideas have been put forward to explain the observed
behavior, and to reconcile the early viewpoint of Wigner with the
possibility of Mott physics in the low density regime. According to
this \emph{Wigner-Mott scenario}\cite{pankov-2008-77,Alberto_NaturePhysics}
the rich phenomenology of the two-dimensional electron gas (2DEG)
is ultimately determined by the competition between the \textit{long-range}
Coulomb interaction and the kinetic energy. As first noted by Wigner,\cite{Wigner34,Giuliani05}
both the the Coulomb interaction $E_{C}$ and the kinetic (Fermi)
energy $E_{F}$ decrease with decreasing density $n$, but their ratio
$r_{s}=E_{C}/E_{F}$, \ie the Wigner-Seitz radius, increases. Thus,
the Coulomb repulsion is expected to dominate at very low densities
($r_{s}\gg1$) and low enough temperatures, leading to the formation
of a charge-ordered state -- a \textit{Wigner}\emph{ crystal.} In
this regime each electron forms a bound state within a potential well
formed by the repulsion from other electrons, and must overcome an
activation gap to escape. In the opposite regime ($r_{s}\ll1$) the
quantum fluctuations lead to the eventual melting of the Wigner lattice,
and the consequent formation of a homogeneous Fermi liquid state.
In between these two limits the system is expected to undergo a series
of non-trivial transformations, making the description of the phase
diagram of the 2DEG one of the most challenging tasks of modern condensed
matter physics.

\subsection{Driving force: interactions or disorder?}
This Wigner-Mott scenario should be contrasted with the alternative
physical picture as first proposed in the early theoretical works
of Refs.~{[}\onlinecite{Finkelstein84,Castellani84}{]}. Here disorder
is envisioned as the primary driving force for electron localization,
and the insulator consists of Anderson-localized electrons bound to
impurities. In this scenario, the primary role of the interactions
is to stabilize the metallic state at higher densities, a possible
mechanism that has been explored in relatively recent theoretical
work.\cite{Punnoose} The key theoretical challenge, therefore, is
to carefully identify the precise consequences of each of these physical
pictures, and assess their respective relevance in light of experiments.
In our work we deliberately ignore all disorder effects, and focus
on investigating the predictions of the Wigner-Mott picture, which
we believe can acount for most qualitative aspects of the puzzling
experimental features. 

To understand how the Mott physics, usually ascribed to half-filled
narrow bands, emerges out of the low-density 2D systems, one should
consider the following argument. A barely melted Wigner crystal is
dominated by the local spin correlations while the short-range Coulomb
repulsion largely precludes double-occupations. Therefore, it becomes
reasonable to view a metal in the vicinity of Wigner crystalization
as a system on the brink of Mott localization. Physically, in the
$r_{s}\gg1$ regime, the Coulomb interaction proves so strong that
it keeps all the electrons at bay, even after the crystal has melted.
The volume fraction available for each electron is thus significantly
reduced, giving rise to the {}``confinement'' of each electron.
This situation is \textit{locally} close to half-filling and the Mott
regime, hence providing an explanation for the (observed) similarities
of a dilute 2D electrons gas with conventional Mott materials characterized
by narrow half-filled bands.

While of plausible significance for diluted 2DEG, the phenomenon of
Wigner-Mott localization applies to many more systems. In fact, based
on the general arguments of Wigner, one may expect that it will emerge
for any partially filled band, where inter site (long-range) Coulomb
interactions are able to induce charge ordering. These effects have
generally been little explored, and were so far mostly studied in
(quasi) one-dimensional systems.\cite{Vojta_ladder-PRB2001,Seo_Merino_charge_ordering-Jpn2006}
However, recent detailed experimental studies on charge ordering phenomena
in layered organic molecular crystals,\cite{Takahashi_COexp-Jpn2006}
and on the cobalt oxide $\mbox{Na}_{x}\mbox{CoO}_{2}$, \cite{Ong_Cava_frustration_Science2004}
have triggered several theoretical investigations of extended Hubbard
models which includes both on-site and nearest neighbor interaction.\cite{Merino-archive2008,McKenzie_charge_order_PRB2001,Motrunich_Lee_tV_PRB2004}
Some aspects of the resulting behavior may be influenced by material-specific
details, such as the form of the lattice. In compounds with triangular
lattice structure, for example, geometrical frustration leads to the
competition of several ground states\cite{Merino_frustrationPRB2006}.
Although still relatively few, all these examples indicate that Wigner-Mott
localization is clearly a very general phenomenon, which is yet to
be explored in detail. Its experimental realizations are often found
in two-dimensional systems, where device geometries (e.g. gating the
2DEG) allows easy control of carrier density, which facilitates accessing
the low density transition region. Still, as a matter of principle,
this phenomenon is not restricted to $d=2$, and thus should be found
in all dimensions. In addition, the charge density wave ordering (CDW)
underpinning can emerge for both long-range inter-site (Coulomb)
interactions, but also for sufficiently strong short range (\eg nearest-neighbor)
repulsion.

To gain insight into its generic features, in this paper we restrict
our attention to the\emph{ simplest model} that illustrates its fundamental
mechanism: Mott localization as driven by charge ordering . This is
accomplished by carefully examining the quarter-filled extended Hubbard
model (EHM), \cite{OnoBullaHewson,merino07prl} which we solve using
single-site dynamical mean field theory\cite{georgesDMFT1996} (DMFT)
methods and a combination of several quantum impurity solvers. The
EHM has already been studied in some detail in previous work by Pietig
et al.\cite{Pietig_PRL1999}, using the DMFT approximation. These
authors found a charge-ordered phase that forms at large values of
the inter-site interaction, by using a combination of non-crossing
approximation (NCA), exact diagonalization (ED), and numerical renormalization
group (NRG) calculations. This work demonstrated the increased importance
of correlation effects in the charge-density wave (CDW) phase, leading
to the formation of a strongly renormalized quasiparticle with the
density of states displaying a narrow peak at the Fermi level. However,
the complete $V$-$U$ phase-diagram for this model has not been determined
and, more importantly, no evidence for the formation of a Wigner-Mott
insulating state has been reported. These and some other interesting
aspects of the 2DEG have been clarified in a more recent short paper,\cite{Alberto_NaturePhysics}
where the DMFT equations for the EHM have been solved at low but finite
temperature using the recently-developed continuous-time Quantum Monte
Carlo algorithm.\cite{Haulectqmc,Werner06} This work identified the
existence of a strongly correlated metallic CDW phase, separating
the homogeneous metal from a Wigner-Mott insulating phase present
for larger values of the non-local interaction and small enough values
of the short-range interaction. In contrast, a direct transition between
the Fermi liquid metal and the Wigner-Mott insulating state was found
at larger values of the local correlation, triggered by the tendency
to charge ordering.

Despite this progress, however, several important physical issues
remain to be clarified, in order to correctly understand the phenomenology
of the Wigner-Mott scenario and its possible relevance for the 2DEG
problem. In particular, even within the considered DMFT solution of
the EHM, the following questions need to be addressed: 
\begin{enumerate}
\item What are the conditions required to find an intermediate strongly
correlated CDW metallic (CDW-M) phase? 
\item What is the temperature dependence of this phase?
\item What is the nature of the phase transitions between the different
phases and how does the critical behavior depend on temperature? 
\item Is there a regime of phase coexistence between the metal and the insulating
state, allowing the possibility of phase separation? 
\end{enumerate}
This paper presents careful calculations providing precise and convincing
answers to all these questions. In particular, we show the existence
of an intermediate CDW-M phase, that separate the homogeneous metal
from the Wigner-Mott insulator, for all temperatures smaller than
a critical value $T=T_{c}$. The shrinking of the CDW-M phase corresponds
to temperature dependent behavior of the Wigner-Mott transition, similar
to the Pomeranchuk effect. \cite{richardson97rmp} We identify two
transitions in the phase-diagram of the model, namely a charge-ordering
transition from the homogeneous metal to CDW-metallic state at $V=V_{c1}(T)$
and a Wigner-Mott MIT taking place at $V=V_{c2}(T)$. We demonstrate
the continuous character of the MIT for $T<T_{c}$ and we give clear
indications for the existence of phase-coexistence, \ie a sharp first-order
transition, at higher temperatures. Furthermore, we present a detailed
study of the evolution of the two critical lines $V_{c1}$ and $V_{c2}$
as a function of increasing temperature from $T=0$. 

The rest of the paper is organized as follows. In section \ref{Sec2}
we introduce the main model and the relative notation, together with
a brief overview of the (numerical) methods used in this work. In
section \ref{Sec3} we present the results concerning the Wigner-Mott
transition in the EHM and we discuss the related phase-diagram. We
conclude in section \ref{Sec4}, where we present our perspective
on the relevance of our picture for the 2DEG and discuss its relation
to alternative theoretical scenarios.

\section{Model and methods}
\label{Sec2}

\subsection{Extended Hubbard model}
\label{Sec2.1} 
The extended Hubbard model is the simplest model that
capture the interplay between strong correlation and charge-ordering
effects. This model includes both a local Coulomb interaction, represented
by the familiar Hubbard $U$ term, and a non-local (inter-site) repulsion
$V$. While the presence of this inter-site repulsion $V$ may induce
the formation of a charge-ordered phase, it also proves capable to
enhance the effectiveness of the on-site repulsion $U$, leading to
the formation of a strongly correlated physics even away from integer-filling.
The lattice structure of the model is introduced in order to capture
the crystalline order of the 2DEG. The lattice constant is then constrained
by the requirement that each cell contains two lattice sites, which
can be regarded as precursors of the interstitial and vacancies in
the Wigner crystal phase, corresponds to an area of $\pi r_{s}^{2}a_{B}^{2}$.
The corresponding extended Hubbard model Hamiltonian takes the form:
\begin{equation}
H=-t\sum_{\langle ij\rangle\sigma}c_{i\sigma}^{\dagger}c_{j\sigma}+U\sum_{i}n_{i\uparrow}n_{i\downarrow}+V\sum_{ij}n_{i}n_{j}-\mu\sum_{i\sigma}n_{i\sigma}.\label{EHM}\end{equation}
Here, $U$ and $V$ are the on-site and the nearest neighbor interactions,
respectively. The parameter $t$ is the hopping amplitude, $c_{i\sigma}^{\dagger}$
($c_{i\sigma}$) are the creation (annihilation) operators and $n_{i}=n_{i\uparrow}+n_{i\downarrow}$
is the occupation number operator on site $i$. The chemical potential
$\mu$ is adjusted to enforce the quarter filling constraint. To this
end we can restrict ourselves to consider a bipartite lattice structure
with sublattices $A$ and $B$.

We solve the problem posed by the Hamiltonian Eq.~\ref{EHM} using
single-site DMFT method,\cite{georgesDMFT1996} which amounts to mapping
the previous lattice problem onto that of a single impurity, coupled
to an effective bath, to be self-consistently determined. This approximation
becomes exact in the infinite coordination number limit,\cite{metzner,metzvol}
provided the hopping parameter and the inter-site interaction are
rescaled as $t\rightarrow t/\sqrt{z}$ and $V\rightarrow2V/z$. In
this limit, the interaction between electrons on neighboring sites
$i$ and $j$ are treated in the Hartree approximation, and the Hamiltonian
Eq.~\ref{EHM} assumes the form: 
\begin{eqnarray}
\begin{split}H= & -t\sum_{<ij>,\sigma}\left(c_{ij,\sigma}c_{ij,\sigma}^{\dagger}+h.c.\right)-\mu\sum_{i}n_{i}+U\sum_{i}n_{i\uparrow}n_{i\downarrow}\\
 & +V\sum_{i\in A,j\in B}\left(\langle n_{i}\rangle n_{j}+n_{i}\langle n_{j}\rangle-\langle n_{i}\rangle\langle n_{j}\rangle\right)
 \end{split}
\label{Hartree}
\end{eqnarray}
 For simplicity, the terms linear in the occupation operators can be easily absorbed in a redefinition of the chemical potential $\mu_{A}=\mu-2V\langle n_{B}\rangle$, $\mu_{B}=\mu-2V\langle n_{A}\rangle$.
The Weiss field, describing the properties of the effective bath in
the DMFT approximation, reads: 
\[G_{0;A,B}^{-1}(i\omega_{n})=i\omega_{n}-\mu_{A,B}-\Delta_{A,B}(i\omega_{n}),\]
where $\Delta_{A,B}(i\omega_{n})$ is the hybridization function
of the associated effective single impurity Anderson model. The self-consistency
condition for the model at hand assume the following expression \cite{georgesDMFT1996}:
\[G_{A,B}=\zeta_{B,A}\int_{\mathbb{R}}d\varepsilon\frac{\rho_{0}(\varepsilon)}{\zeta_{A}\zeta_{B}-\varepsilon^{2}}\]
 with $\zeta_{A,B}=i\omega_{n}+\mu_{A,B}-\Sigma_{A,B}(i\omega_{n})$,
$\Sigma_{A,B}$ being the \textit{local} self-energy function and
$\rho_{0}(\varepsilon)=\sum_{{\bf k}}\delta(\varepsilon-\varepsilon({\bf k}))$
the non-interacting density of states corresponding to the chosen
lattice. In the following we focus on the simple semi-circular model
density of states, for which $\rho_{0}(\varepsilon)=\sqrt{D^{2}-\varepsilon^{2}}/4\pi t^{2}$
with $D=2t=1$ fixing the energy units of the problem. The self-consistency
equations then reduce to: 
\begin{equation}
\Delta_{A,B}=t^{2}G_{B,A}\,.
\label{SC}\end{equation}

\subsection{Methods}
\label{Sec2.3} 
In this section we briefly review the methods of solutions used in
this work, emphasizing the advantages and the drawbacks for each of
them. More detailed review of the same methods can be found elsewhere
in the literature, cf. Ref.~\onlinecite{georgesDMFT1996,Haulectqmc,daniel1}.
We begin our discussion with the simplest used impurity solver, namely
the four slave-boson (SB4) method of Kotliar and Ruckenstein. \cite{kotliarsb}
At the mean field level (saddle-point), this method at $T=0$ proves
to be equivalent to the well-known Gutzwiller variational approximation,
but it also allows for extensions to $T>0$. Within the SB4 method
the effective impurity problem is reduced to the solution of a set
of algebraic equations for the appropriate slave-boson variational
parameters. Solving these equations is a relatively simple task, which
allows for quite an accurate numerical solution, and a very precise
characterization of the leading critical behavior. Although the SB4
method variationally calculates the Fermi liquid parameters specifying
the quasiparticles and by neglecting incoherent parts of the spectrum,
this method is known to generally provide qualitatively and largely
even quantitatively accurate solutions when applied within the framework
of DMFT theories. In the SB4 approach, the low energy part of the
local Green's function is parametrized in terms of the quasiparticle
weight $Z_{i}$ at site $i$. In the CDW phase these quantities differ
on the two sublattices $i=A,B$. The quasiparticle weight $Z_{i}[e_{i},d_{i}]$
is expressed through the mean field slave-boson parameters $e_{i}^{2}$
and $d_{i}^{2}$, which are equal to the probability that the site
is empty and doubly occupied, respectively. These parameters and ultimately
the impurity Green's functions $G_{fA,fB}$, for either the $A$ and
the $B$ sublattices, are determined by constrained minimization of
a suitable local free energy ${\cal F}_{i}$ parametrized by a Lagrange's
multiplier $\lambda_{i}$: 
\begin{align}
{\cal F}_{i}= & -\frac{2}{\beta}\sum_{\omega_{n}}\ln[-i\omega_{n}-\tilde{\mu_{i}}+Z_{i}\Delta_{i}(i\omega_{n})]+Ud_{i}^{2}\nonumber \\
 & -\lambda_{i}(1-e_{i}^{2}+d_{i}^{2})
 \end{align}
 with $\tilde{\mu}_{A,B}=\mu-2Vn_{B,A}-\lambda_{A,B}$. The (low energy
part of the) local Green's functions are determined by the relation
$G_{A,B}=Z_{A,B}G_{fA,fB}$. The DMFT equations are closed by supplementig
these relation with the self-consistency condition \ref{SC}. The
chemical potential $\mu$ and occupation numbers $n_{A}$ and $n_{B}$
also have to be calculated self-consistently in order to enforce quarter-filling
constraint $n_{A}+n_{B}=1$, where $n_{A,B}=\frac{2}{\beta}\sum_{\omega_{n}}G_{fA,B}(i\omega_{n})$.

To supplement the SB4 results with more accurate methods we solved
the DMFT equation for the EHM with (numerically) exact methods. In
particular in the $T=0$ limit we implemented a Density Matrix Renormalization
Group (DMRG) impurity solver. The DMRG idea have first been introduced
in $1992$ by S.~White \cite{White92} to deal with one-dimensional
quantum lattice problems,\cite{dmrgRMP} for which the Wilson's real-space
blocking scheme was observed to fail in describing the correct solution.
In more recent work,\cite{daniel1,daniel2} DMRG has been adapted
to solve the self-consistent DMFT equations. The algorithm is based
on a recursive exact diagonalization (ED) of the associated quantum
impurity problem with an increasing size of the effective bath. Similarly
to other renormalization group based method, the DMRG provides a suitable
method to restrict the solution of the effective problem to its relevant
subspace. This is achieved using a clever method for limiting the
exponential growth of the Hilbert space dimension, based on the analysis
of the reduced density matrix. Starting with a small effective bath
in the form of two linear chains,\cite{georgesDMFT1996} the effective
impurity problem is exactly solved, and then the size of the bath
is recursively increased. At each step the Hilbert space of the problem
is constructed using the basis formed by the first $M$ (\ie the
{}``most probable'') eigenvectors of the reduced density matrix
$\hat{\rho}=\mbox{Tr}_{\lvert_{Env}}|gs\rangle\langle gs|$, where
$|gs\rangle$ is the (approximated) ground state and the environment
is represented by the impurity plus one of the two bath' chains. For
every fixed size of the impurity problem, the impurity Green's function
are evaluated using successive applications of the Lanczos' method.\cite{karenPRB}
The self-consistency condition \ref{SC} is then used to update the
Hamiltonian parameters of the problem.

The recursive nature of the DMRG method permits a systematic improvement
of the quality of the solution. In addition, this method has the advantage
of treating on the same footing both low- and high-energy scales,
in contrast to standard RG technique, which principally focus on the
low-energy spectra of the problem. However, the particular one-dimensional
topology imposed by the DMRG algorithm makes this method suffering
of the finite size effects. In practice effective baths as large as
$L=20$ sites have to be reached in order to obtain a satisfying solution
of the model.

While the DMRG algorithm it is straightforward at zero temperature,
its extensions to finite temperature regime are non-trivial. To this
end, we have solved the DMFT equations using alternative algorithms.
The first finite temperature method is (complete) Exact Diagonalization
of the discretized effective impurity problem. This method permits
to access the full spectrum of the problem, thus allowing for the
calculation of the exact impurity Green's function and other observables.
The Hamiltonian parameters are self-consistently determined using
an adaptive method that permits to minimize the (in principle large)
finite size effects.\cite{caffarel_edstar} This method is indeed
known to have an almost constant scaling with the bath' size. On the
other hand, a small number of degrees of freedom in the problem prevents
the access to the very low energy physics of the model, thus a significant
number of bath' sites ($N_{s}>5$) becomes necessary to have a good
description of the model solution. Our calculation has been performed
using $N_{s}=7$, which is the largest accessible number of sites
for this method.

The second method we used to solve the DMFT equations at finite temperature
is the Continuous Time Quantum Monte Carlo, in the implementation
of Ref.~{[}\onlinecite{Haulectqmc}{]}. The CTQMC is a statistically
exact method based on the Monte Carlo samplings of the diagrams obtained
by perturbative expansion of the impurity problem with respect to
the hybridization. The CTQMC algorithm of Ref.~{[}\onlinecite{Haulectqmc}{]}
has proven to be a highly reliable and stable method, that permit
to capture the low temperature regime of the model. On the other hand,
the statistical nature of the method put some, indeed non-stringent,
limitations to its applicability in the proximity of a phase transition.
In this regime, the solution requires a very large number of Monte
Carlo samplings and a large number of iterations of the self-consistency
algorithm to overcome the enhanced fluctuations and the critical slowing
down of the solution.

\subsection{The $U=0$ solution}

\begin{figure}
\centering \includegraphics[width=1\linewidth]{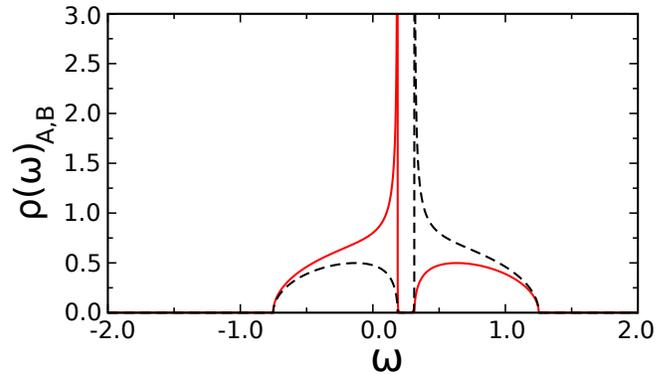} 
\caption{(Color online) Non-interacting DOS of sublattice $A$ (solid line)
and $B$ (dashed line) for $\delta\simeq0.2$. The system is in a
CDW-M state and correspondingly the DOS is split in two lobes. The
Fermi energy, corresponding to $\omega=0$, falls inside the lower
lobe as results of the quarter filling constraint.}
\label{U0solution} 
\end{figure}

In the non-interacting limit $U=0$ the problem posed by the Hamiltonian
Eq.~\ref{Hartree} can be easily solved analytically. In particular
here we are interested in obtaining the analytic solution corresponding
to the DMFT approximation, in order to get some physical insight from
the model. In the non-interacting limit the DMFT equations can be
recast in the form: 
\begin{equation}
\begin{split} & G_{A}^{-1}(z)=z+\mu_{A}-t^{2}G_{B}=\alpha-t^{2}G_{B}\\
 & G_{B}^{-1}(z)=z+\mu_{B}-t^{2}G_{A}=\beta-t^{2}G_{A}\end{split}
\end{equation}
 and solved for, say, $G_{A}$ giving: 
 \[G_{A}=2\left[\beta\pm\sqrt{\beta^{2}-\beta/\alpha}\right]\]
 All the solutions to this equation can be expressed in terms of a
single parameter $\delta=n_{A}-n_{B}$ representing the occupation
imbalance between the two sublattices. The density of states is non-zero
only in the energy interval defined by the $\beta^{2}-\beta/\alpha<0$
condition. The system undergoes a charge ordering at $V=V_{c}$ with
the formation of a spectral gap $\Delta=V\delta$ (for $V\geq V_{c}$).
In this regime the original quarter-filled band splits into two bands,
but the Fermi level remains inside the lowest band, which remains
half-filled, cf. Fig.~\ref{U0solution}. Thus a spectral gap at the
Fermi level can only be opened by increasing the local correlation
$U$.

In this limit we can think at the system as consisting of one sublattice
$A$ with occupation close to half filling, and the other $B$ as
being nearly empty. An estimate of the effective hopping between sites
in the sublattice $A$ can be obtained integrating out (virtual) high
energy processes, corresponding to hops through the sites of sublattice
$B$. At $V\gg t$, to leading order the effective hopping between
the sites of the $A$ sublattice then takes the form $t_{eff}=t^{2}/V$.
\begin{figure}
\centering \includegraphics[clip,width=1\linewidth]{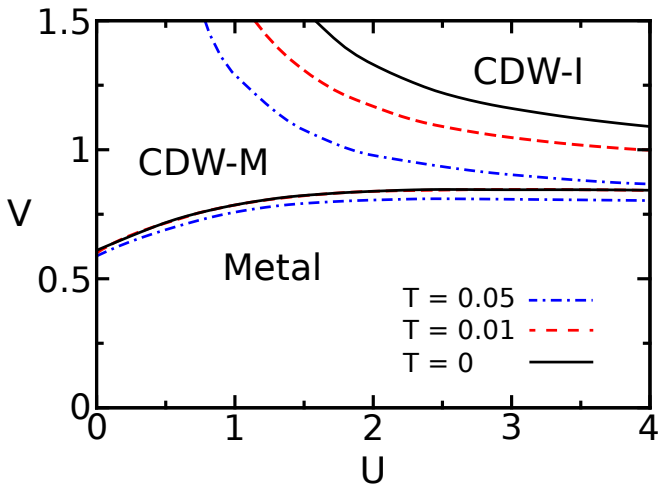} 
\caption{(Color online) DMFT phase diagram of the extended Hubbard model obtained
with a slave-boson (SB4) impurity solver. The figure shows the phase-boundaries
obtained for different values of the temperature $T$. At $T=0$ (solid
line) the solution shows the existence of a large CDW metallic phase
at all values of the local correlation $U$. This region narrows as
the temperature T is increased from zero to $T=0.01$ (dashed line)
to $T=0.05$ (dot-dashed line), but it never disappears.}
\label{PDsb4} 
\end{figure}

\section{The Wigner-Mott transition}
\label{Sec3}

\subsection{Phase-diagram}
In this section we present results concerning the phase-diagram of
the EHM as a function of the local correlation $U$ and the inter-site
interaction $V$. In the following we use different methods to investigate
the various phases of the system as a function of the temperature
$T$.

Our results show the existence of three qualitatively different phases,
which can be classified by two order parameters: i) the CDW staggered
density $\delta=n_{A}-n_{B}$ and ii) the quasi-particle residue (\ie renormalization
constant) $Z_{A,B}$ at each sublattice. The different phases of the
model are characterized as following: 
\begin{itemize}
\item \textbf{Fermi liquid}. A featureless homogeneous metallic state, corresponding
to $n_{A}=n_{B}=1/2$ and $Z_{A}=Z_{B}>0$. 
\item \textbf{CDW-M}. A CDW-ordered metallic state with $n_{A}>n_{B}$ and
$Z_{B}>Z_{A}$. The sublattice $A$ has an occupation near to the
half-filling condition ($n_{A}\simeq1$), thus closer to Mott localization.
As a consequence of this the quasi-particles weight $Z_{A}$ is substantially
reduced, corresponding to a enhancement of their effect mass $m_{A}^{*}/m_{A}\simeq Z_{A}^{-1}$. 
\item \textbf{CDW-I}. A Wigner-Mott insulating phase. In this regime $n_{A}\gg n_{B}$
but while $Z_{B}\sim1$, sublattice $A$ has an identically zero quasi-particle
weight $Z_{A}=0$. This corresponds to the Mott localization of the
electrons on the {}``nearly half-filled'' sublattice $A$. 
\end{itemize}
\begin{figure}
\includegraphics[clip,width=1\linewidth]{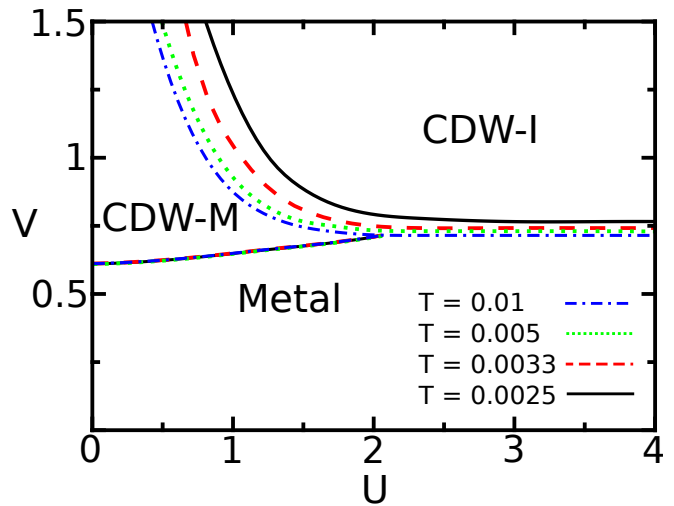} 
\caption{(Color online) DMFT phase diagram of the extended Hubbard model obtained
with CTQMC method. The results illustrate the evolution of the phase-boundaries
for the Mott-Wigner transition as a function of increasing temperature.
For the lowest value of the temperature $T=0.0025$ (solid line) we
observe the well established presence of a CDW-M separating the homogeneous
metal from the CDW-insulator. This phase-region narrows upon increasing
the value of the temperature to $T=0.0033$ (dashed line) and $T=0.005$
(dotted line), till a closure of the phase-boundaries is observed
for $T\sim0.01$ (dot-dashed line) at $U\sim2$.}
\label{PDctqmc} 
\end{figure}

The $U-V$ phase diagram as obtained from our SB4 solution for increasing
values of the temperature $T$ is presented in Fig.~\ref{PDsb4}.
At zero temperature, we obtain a continuous transition from the homogeneous
metallic phase to the CDW-metal. This is followed by a second continuous
transition to the CDW-insulator as the inter-site interaction $V$
is further increased. The intermediate CDW metallic phase, while decreasing
in size at large $U$, remains of finite extent even in the limit
$U\rightarrow\infty$ limit. Upon increasing the temperature $T$,
we observe a shrinking of the intermediate CDW-M phase at large values
of $U$. Still, no signs for the disappearance of this phase have
been observed with the slave-bosons methods. This result is in contrast
to that initially obtained in Ref.~\onlinecite{Alberto_NaturePhysics}
at very low but finite temperature $T=0.01$ using a more sophisticated
method. Despite the qualitative validity of the SB4 method, this result
clearly puts in question the main features of the phase-diagram for
the EHM. In particular, it suggests that the intermediate CDW-M phase
at intermediate to large U is very fragile to temperature. This observation
calls for a more detailed study of the temperature dependence of the
metal-insulator phase boundary, which we examine in the following
sections.

\begin{figure}
\includegraphics[clip,width=1\linewidth]{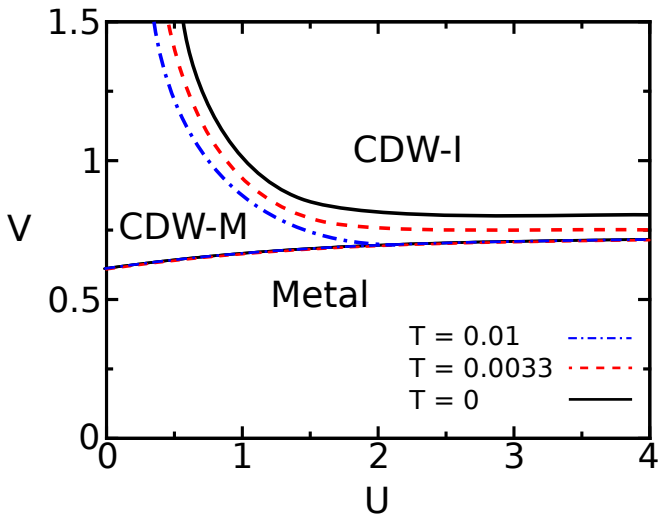} 
\caption{(Color online) DMFT phase diagram of the extended Hubbard model obtained
with DMRG and Exact Diagonalization (finite temperature). The figure
shows the evolution of the phase-boundaries as a function of increasing
temperature from $T=0$ (solid line) to $T=0.0033$ (dashed line),
to $T=0.01$ (dot-dashed line). The intermediate CDW-M observed at
$T=0$ is found to narrows upon increasing the temperature, until
a closure of the boundary-lines is obtained around $U\sim2$ and for
$T\sim0.01$. The DMRG results have been obtained with a bath of $L=30$
sites, whereas the ED calculation have been performed with a number
of sites $N=7$.}
\label{PDdmrged} 
\end{figure}

The results from CTQMC solution of the DMFT equations for the EHM
are summarized in the phase-diagram in Fig.~\ref{PDctqmc}. While
this method cannot capture the zero temperature properties of the
model, it provides an excellent and reliable solution at very-low
temperature scales. At the lowest accessible temperature $T=0.0025$
we observe a continuous transition between a homogeneous metallic
state to a CDW-metal for low values of $V$ and every value of $U$.
At this value of the temperature the CDW-M phase is found to be destabilized
towards a CDW-insulating phase upon increasing the inter-site interaction
$V$. Upon increasing the temperature we observed a narrowing of the
intermediate CDW-M in favor of the Wigner-Mott insulating state, until
a merge of the boundary lines is observed for temperatures of the
order $T\sim0.01$ (that is two orders of magnitude smaller than the
bare bandwidth).

To better clarify the issue of closing boundary-lines at low temperature
we have solved the DMFT equations with an exact method at zero temperature,
\ie the DMRG. Furthermore, we have complemented this investigation
with a powerful ED solver, in order to access the finite temperature
properties of the system. The ED-DMRG results for the extended Hubbard
model problem are condensed in the $U-V$ phase diagram presented
in Fig.~\ref{PDdmrged}. The phase-diagram is in very good agreement
with that obtained from the CTQMC solution of the model, already presented
in Fig.~\ref{PDctqmc}. Our calculations confirm the existence of
an homogeneous metallic phase at low values of $V$ and any value
of the local correlation $U$. The stability of the homogeneous metallic
phase is related to the quarter filling condition, making the local
interaction nearly ineffective at small $V$. For any fixed value
of $U$ and for $V$ larger than a critical value $V_{c1}$ almost
independent of the temperature $T$, the system shows a continuous
phase-transition towards a CDW-M state. The evolution of the two sublattices
upon further increasing $V$ is different. One sublattice gets nearly
empty and becomes, for sufficiently large U, a band insulator. The
other sublattice, with a filling closer to $1$, undergoes a continuous
Mott transition towards a CDW-Insulating state for $U$ large enough
and $V$ larger than a critical value $V_{c2}$ weakly depending on
the temperature $T$. In particular the zero temperature solution
of the model shows the persistence of the intermediate CDW-M for any
value of the local correlation $U$. This phase is observed to narrow
upon increasing the temperature until a closure of the boundary lines
at $U\sim2$ is obtained for $T\sim0.01$, in excellent agreement
with the CTQMC solution of the model.

\begin{figure}
\centering \includegraphics[clip,width=1\linewidth]{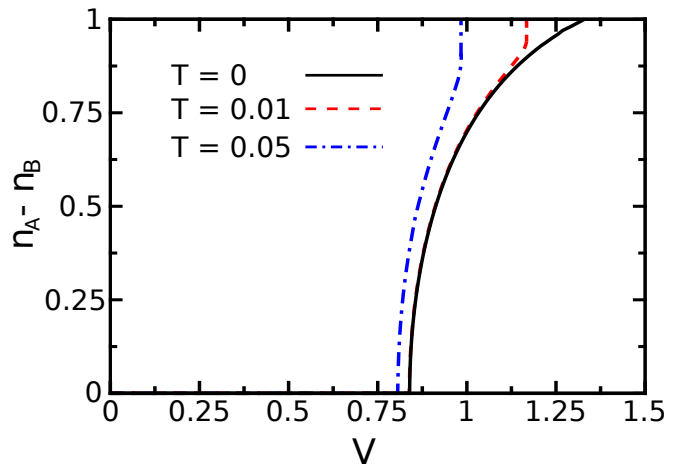} 
\caption{(Color online) Order parameter $\delta=n_{A}-n_{B}$ as a function
of the inter-site interaction $V$, for $U=2$ and increasing values
of the temperature $T$. The metallic CDW phase appears for $V>V_{c1}\simeq0.84$,
while a CDW-insulator forms for $V>V_{c2}\simeq1.35$. Within SB4
solution, the $T=0$ transition is continuous, while for $T>0$ a
small discontinuity in CDW order parameter is observed, that is a
possible artifact of the method.}
\label{deltanSB4} 
\end{figure}

\subsection{Generalized Pomeranchuk effect}
Next, we would like to discuss the physical implications of the observed
temperature evolution of the phase diagram, in relation to the interesting
non-monotonic resistivity behavior observed in very clean 2DEG samples.\cite{kravchenko1994}
As a general trend we observed a shrinking of the CDW-M region as
temperature is increased from $T=0$. This effect is most pronounced
at the CDW-M to CDW-I boundary $V_{c2}(U)$, i.e. in the region where
a heavy Fermi liquid forms. From the physical point of view, this
temperature dependent behavior reflects the entropy gained in destroying
the (heavy) Fermi liquid to form localized magnetic moments in the
Wigner-Mott (CDW-I) phase. This mechanism of entropy release is indeed
similar to that anticipated in the early work of Pomeranchuk,\cite{pomeranchuk50ZETP}
who speculated about the general problem finite-temperature solidification
of $^{3}$He.\cite{richardson97rmp}

\begin{figure}
\subfigure[\, DMRG]{\label{delta2} \includegraphics[clip,width=1\linewidth]{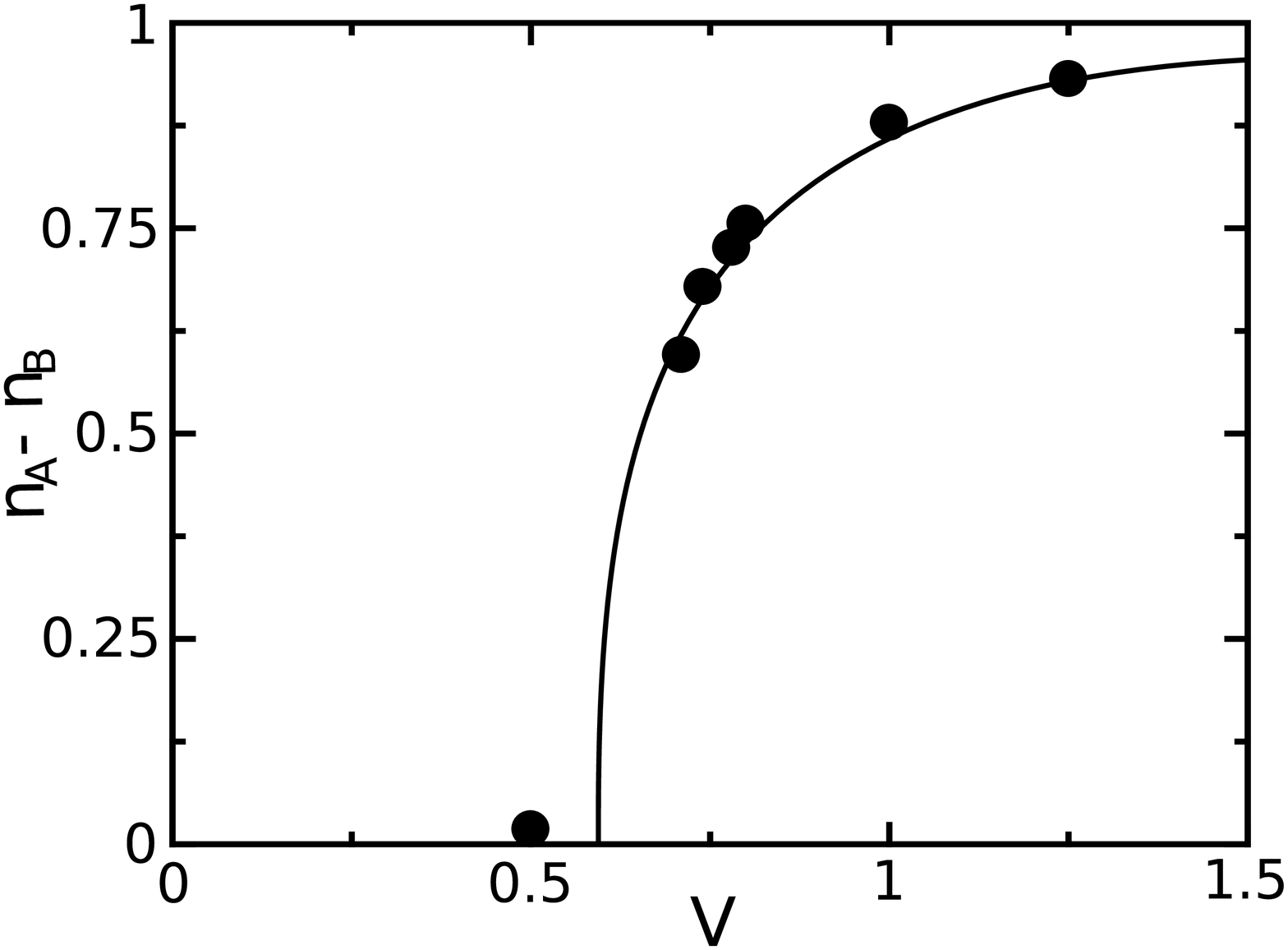}}
\subfigure[\, CTQMC]{\label{delta1} \includegraphics[clip,width=1\linewidth]{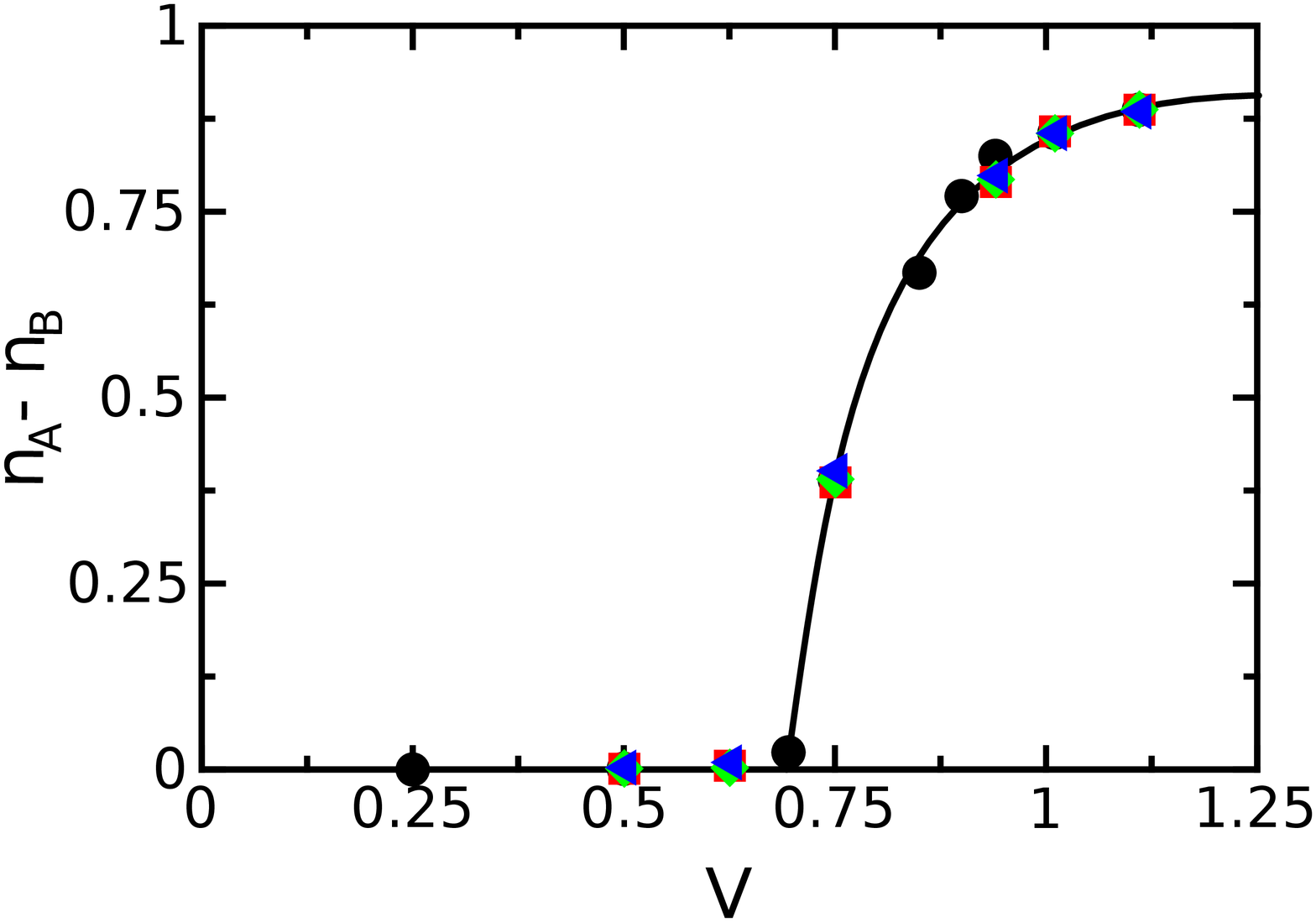}} 
\caption{(Color online) The two panels show the behavior of the order parameter
$\delta=n_{A}-n_{B}$ as a function of inter-site interaction $V$,
for $U=2$ and different values of the temperature $T$. (a) Results
from zero temperature calculations with DMRG method. (b) CTQMC solution
for several temperatures $T=0.01$ (black dots), $0.005$ (red squares),
$0.0033$ (green diamond), $0.0025$ (blue triangles). Both $T=0$
and finite temperature calculations display a rather conventional
$(V-V_{c1})^{1/2}$ critical behavior near the CDW transition, similar
to that observed in the non-interacting limit.}
\end{figure}

It should be emphasized, however, that this {}``entropic'' destruction
of a strongly correlated Fermi liquid is a very general phenomenon,
which does not necessarily require a first order phase transition,
as postulated by Pomeranchuk, or a phase coexistence with separation
of the phases.\cite{spivak05prl,spivak04prb,spivak03prb,spivak01prb}
For example, the same effect can be observed in many heavy fermion
systems, where the destruction of the heavy Fermi liquid for temperatures
larger than the {}``coherence'' temperature $T^{*}\sim T_{K}$,
coincides with a release of extensive spin entropy $S\sim k_{B}\ln2$,
corresponding to the localization of the $f$-electrons. The corrrsponding
resistivity maximum represents a temperature-driven crossover rather
than a sharp a phase-transition, related to the fact that the two
phases do not have to differ by symmetry. In the Wigner-Mott picture
of the 2DEG, we thus recognize in the entropy of the high temperature
state as the driving force for electron localization.

\subsection{Charge ordering}
To better understand the nature of the continuous transitions observed
in the phase-diagrams of Sec.~\ref{Sec2}, we now investigate the
behavior of the charge order parameters $\delta=n_{A}-n_{B}$ as a
function of $V$ for different values of the temperature. The symmetry
breaking associated to this transition corresponds to the occupation
unbalance in favor of one of the two sublattices and the order parameter
$\delta$ roughly measures the tendency of the system to form charge
ordered phase. All the following results have been obtained for a
fixed value of the local correlation $U=2$, corresponding to a region
of the phase-diagram with narrowing intermediate CDW-M phase separating
the Wigner-Mott insulator from the homogeneous metal. 
\begin{figure}
\centering \includegraphics[clip,width=1\linewidth]{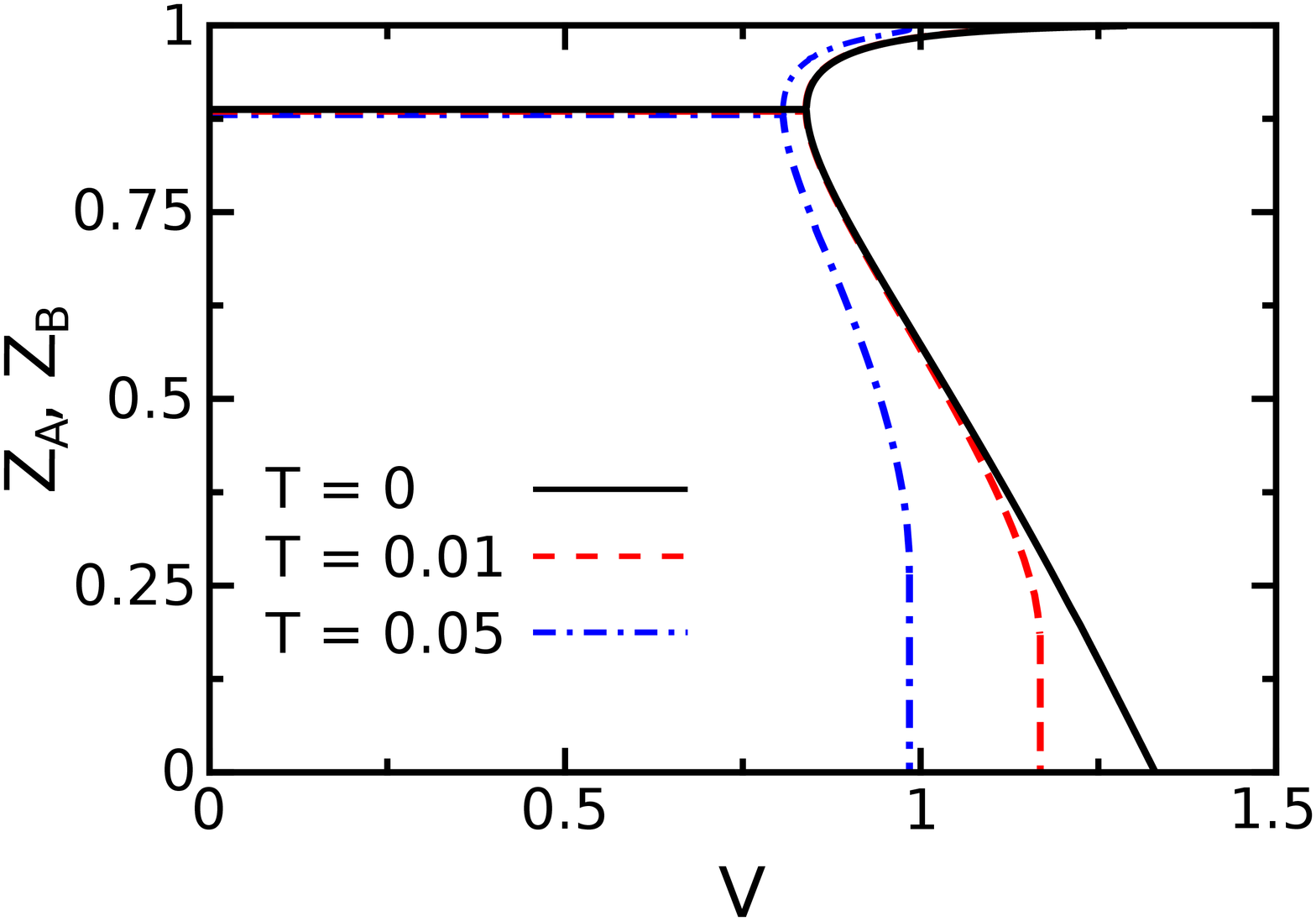} 
\caption{(Color online) Quasi-particle weight $Z_{A,B}$ as a function of $V$
for different temperatures and $U=2$. At $T=0$ (solid line) we observe
a linear vanishing of the non-empty sublattice renormalization constant,
while the constant corresponding to nearly-empty sublattice flows
to one. At larger value of the temperature $T=0.01$ (dot-dashed line)
and $T=0.05$ (dashed line) the SB4 predicts a small jump in the quasi-particle
weight, that is a possible artifact of the method rather than a real
discontinuity of the transition.}
\label{ZabSB4} 
\end{figure}

\begin{figure}
\subfigure[\, CTQMC]{\label{Zab1} \includegraphics[clip,width=1\linewidth]{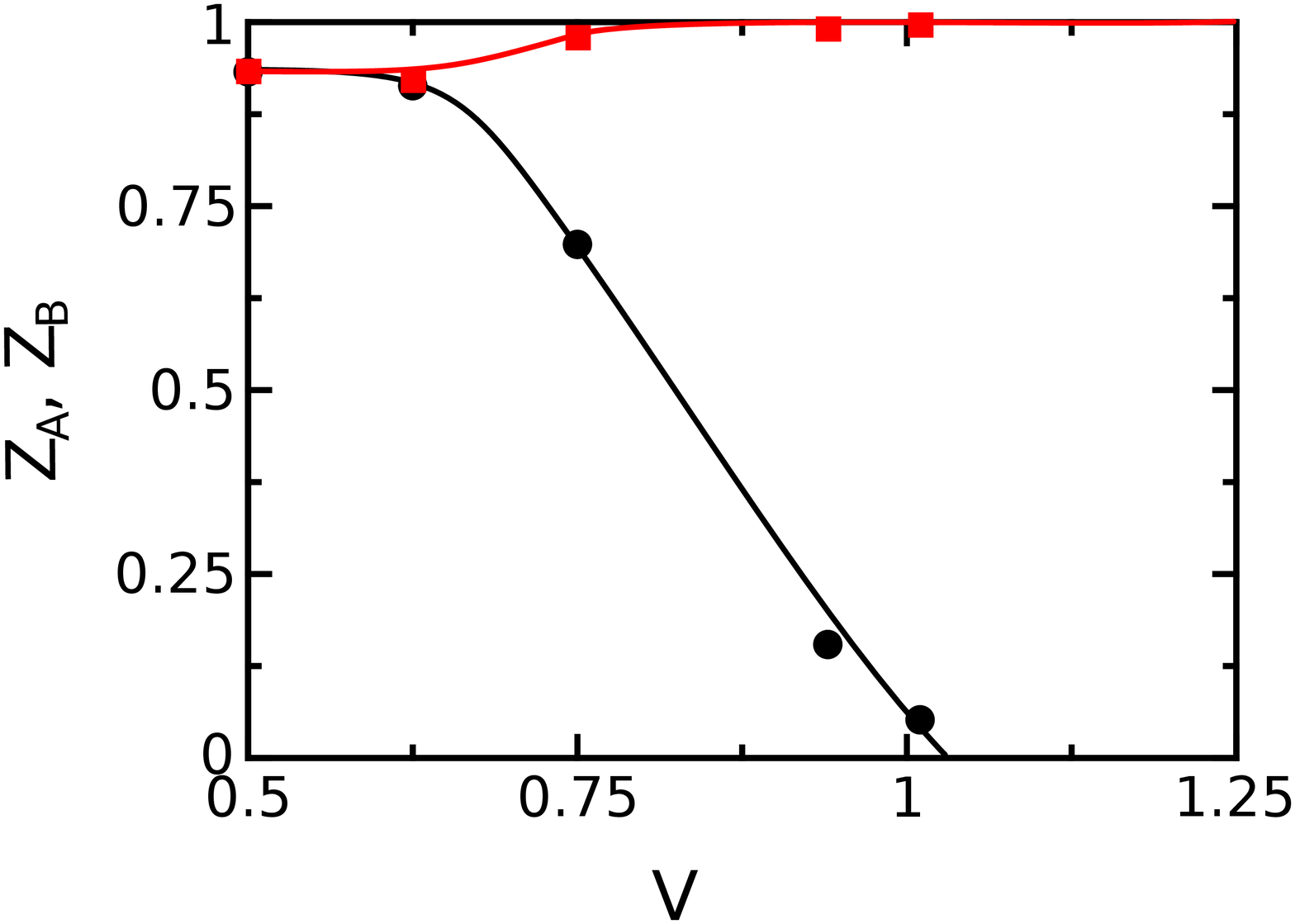}}
\subfigure[\, DMRG]{\label{Zab2} \includegraphics[clip,width=0.9\linewidth]{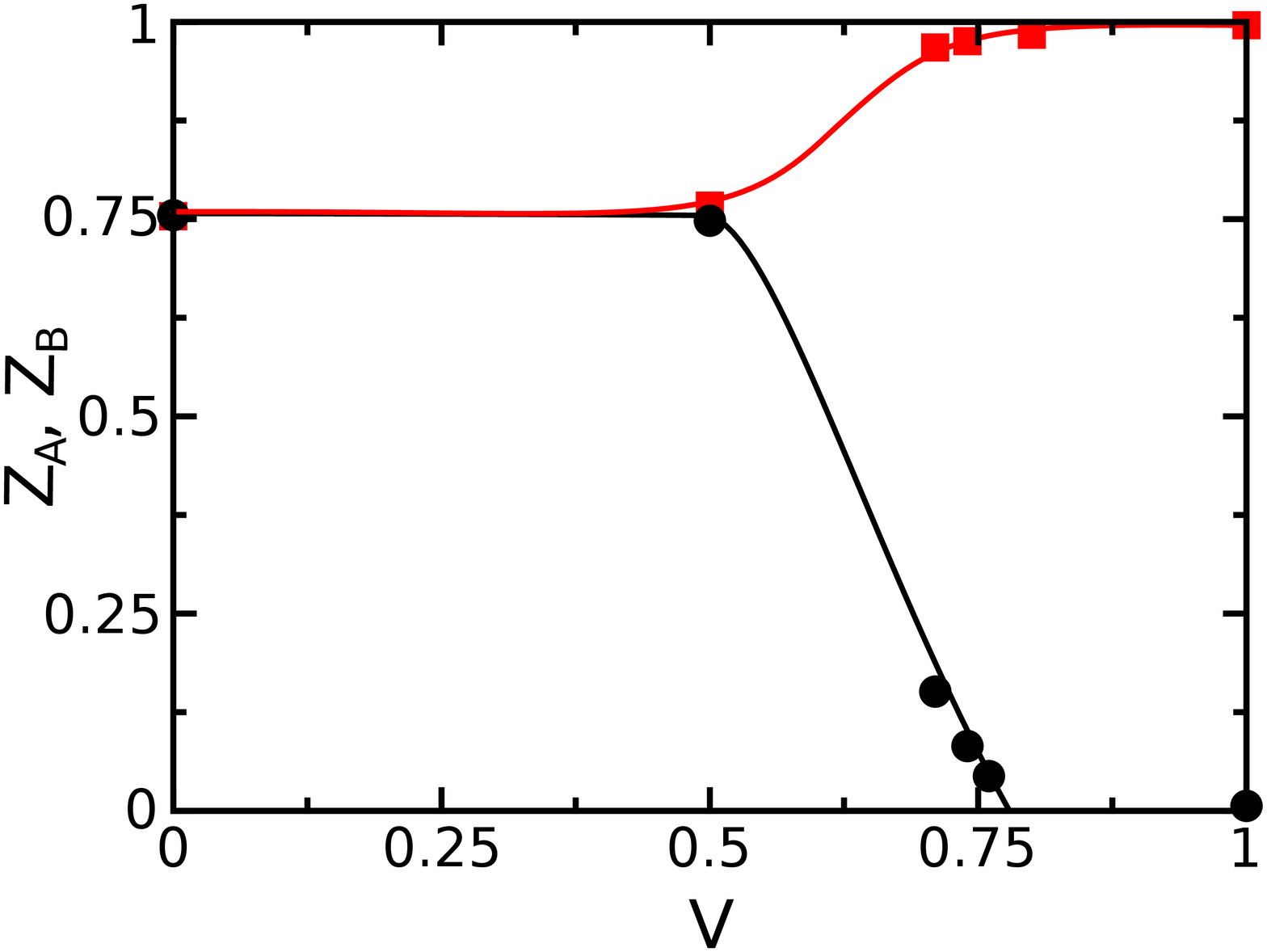}} 
\caption{(Color online) Quasi-particle residues $Z_{A,B}$ behavior as a function
of $V$ and $U=2$ at (a) $T=0.01$ and (b) zero temperature. The
figures show the linear vanishing of the order parameter of the Wigner-Mott
transition, independent of the value of the temperature and expressing
the continous character of the transition.}
\end{figure}

Results from SB4 method are presented in Fig.~\ref{deltanSB4}. The
figure shows, at $V=V_{c1}$, the emergence of a charge ordering instability
which emerges before the Fermi liquid state is destroyed. At $T=0$
the slave-bosons solution displays the conventional critical square-root
behavior for the order parameter, i.e. $\delta=(V-V_{c1})^{1/2}$.
At finite temperature, however, a small {}``jump'' in the order
parameter can be observed, that is believed to be an artifact of the
SB4 method. In the following we shall clarify the character of the
transition using numerically exact methods, namely the CTQMC and DMRG
at finite and zero temperature, respectively.

The DMRG results for the charge ordering at zero temperature are shown
in Fig.~\ref{delta2}. We observe a sharp increasing of the order
parameter for $V\sim V_{c1}$ with a typical square-root-like critical
behavior for higher values of the inter-site correlation. This result
is in good agreement with that obtained with SB4 method at $T=0$,
although the finite size of the DMRG effective problem does not permit
to exactly locate the critical value of $V$. On the other hand, no
trace of any discontinuoities for the order parameter has been observed
at finite temperature, using CTQMC (cf. Fig.~\ref{delta1}) and finite
temperature Exact Diagonalization method (not shown). Thus, our results
put strong evidences for the continuous character of the charge-ordering
transition, irrespective of the value of the temperature.

\subsection{Critical behavior}
Finally we have investigated the destruction of the charge-ordered metallic
state in favor of the Wigner-Mott insulator, driven by the increasing
inter-site correlation $V$ and for large values of the local interaction
$U$. As $V$ increases, the effective hopping amplitude at the non-empty
sublattice $t_{eff}=t^{2}/V$ decreases. Thus, it is reasonable to
expect that for $U\sim t_{eff}$ a Wigner-Mott transition takes place.
Our results, obtained with different methods, substantiate this qualitative
picture. We observe in fact a Wigner-Mott transition for $V_{c2}(U)\sim t_{eff}^{2}/U$,
that is compatible with the observed $1/U$ behavior for the CDW metal
to insulator transition line in the phase-diagram (cf. Sec.~\ref{Sec2}).
In the following we present a characterization of the Wigner-Mott
transition in the EHM in terms of the vanishing of the renormalizaton
constants $Z_{A,B}$. 
\begin{figure}
\includegraphics[clip,width=1\linewidth]{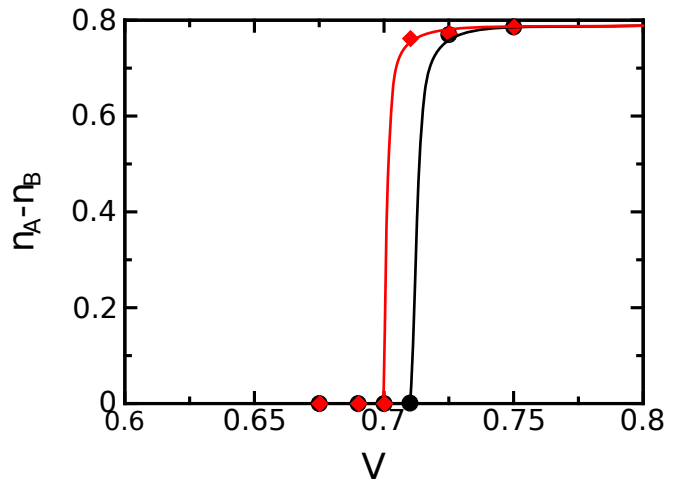} 
\caption{(Color online) Order parameter $\delta=n_{A}-n_{B}$ as a function
of the non-local interaction $V$. Data are obtain from CTQMC calculations
at $T=0.01$ and $U=3$. The figure illustrates the first order-character
of the direct transition between the homogeneous metal and the Wigner-Mott
insulator, presenting evidences for the coexistence of the metallic
and insulating phases in a narrow interval around $V_{c2}$.}
\label{Coex} 
\end{figure}

Results obtained with SB4 method are presented in figure Fig.~\ref{ZabSB4}.
At $T=0$ the quasiparticle weight for the nearly half-filled sublattice,
say $A$, is found to have a linearly vanishing behavior $Z_{A}\simeq(V_{c2}-V)$,
similar to that observed in the conventional Mott scenario. At finite
temperature the SB4 shows again the breakdown of the solution, leading
to unphysical jump in the renormalization constant behavior. This
issue has been clarified using the more sophisticated CTQMC method
at finite temperature. Results from CTQMC calculations at low temperature
are shown in figure Fig.~\ref{Zab1}. The renormalization constant
for the non-empty sublattice is found to have a linear vanishing behavior
as observed within the SB4 solution at $T=0$, thus indicating the
continuous character of the Wigner-Mott transition. A very similar
behavior is observed in the zero temperature model calculations performed
with DMRG and presented in Fig.~\ref{Zab2}.

The continuous transition from the correlated CDW-Metallic phase with
$m^{*}>1$ to the Mott-Wigner insulator, observed at temperatures
$T<T_{c}\simeq0.01$ and local interaction large enough $U>2$, is
replaced at higher temperature $T>T_{c}$ by a sharp first-order transition
from the homogeneous metal with $m^{*}\simeq1$ directly to the CDW-Insulating
state. This is compatible with the shrinking and disappearance of
the intermediate correlated CDW-Metallic state with increasing temperature.
This effect is well illustrated in Fig.~\ref{Coex}, presenting the
tiny hysteresis cycle of the order parameter $\delta=n_{A}-n_{B}$
for a large value of the correlation $U=3$ and $T=0.01$. The two
curves in the figures are obtained following the metallic and the
insulating solutions, respectively. As we can clearly see, the two
solutions have a small region of coexistence around the transition
point at $V=V_{c2}$, indicating the first-order character of the
Wigner-Mott transition at $T>T_{c}$.

Interestingly, no trace of first-order transition have been found
in the less correlated regime $U<2$, which is more relevant for the
interpretation of the experimental results. In this regime the correlated
quasiparticles are gradually destroyed by the thermal fluctuations
at $T>T^{*}\sim Z$, leading to a thermal metal to insulator crossover
with an associated resistivity maximum, but without phase separation
or any first-order transition,\cite{spivak04prb} compatible with
out interpretation in terms of generalized Pomeranchuk effect.

\section{Conclusions and Perspectives}
\label{Sec4} 
In this work we investigated the quarter filled extended
Hubbard model, in order to describe the essential features of the
Wigner-Mott metal-insulator transitions. Using single-site DMFT theory,
we obtained the $U$-$V$ phase-diagram, and carefully studied its
evolution as a function of temperature. For this model, a metallic
charge ordered phase (CDW-M) is generally found at low temperatures,
separating a Wigner-Mott insulator at strong coupling from a homogeneous
Fermi liquid at weak coupling. This intermediate CDW-M phase is the
one showing the most interesting features, chiefly the emergence of
strong correlation effects signaled by large effective mass $m^{*}$
enhancements. Physically, this reflects the presence of heavy quasiparticles
existing only below a characteristic energy scale $E^{*}\sim T_{F}/m^{*}$,
which vanishes at the metal-insulator transition. Indeed, we predict
that such a correlated metallic state can easily be suppressed either
by increasing the temperature beyond a modest temperature $T^{*}\sim E^{*}$,
or by applying modest polarization fields $H^{*}\sim T^{*}$ - in
striking agreement with the experiments\cite{kravchenko1994}. In
addition, we demonstrated that the region occupied by the CDW-M phase
shrinks as a function of increasing temperature, thus preducing an
interesting Pomeranchuk-like effect. We presented an interpretation
of this effect in terms of the entropic destruction of the strongly
renormalized Fermi liquid (CDW-M) in favor the Wigner-Mott insulator,
similar to the early idea proposed by Pomeranchuk in the context of
the $^{3}$He solidification.

The extended Hubbard model we used certainly cannot be regarded as
a realistic or quantitatively accurate represetation of the 2DEG materials.
Ours is an approach complementary to that provided by first principle
(e.g. diffusion Monte Carlo) studies of realistic models\cite{waintal10prb}
of 2DEG -- yet with surprisingly similar results. Both approaches
portray a picture of a strongly correlated electron fluid featuring
a single characteristic energy scale. But precisely by virtue of its
simplicity, our model calculation makes it possible to unravel\emph{
the mechanism} behind the puzzling behavior in the metal-insulator
transition region. It demonstrates how the tendency to charge ordering
reinforces the transmutation of conduction electrons into local magnetic
moments -- a fundamental physical process behind the phenomenon Wigner-Mott
localization. 

Many quantitative aspects of our model can and should be improved.
In particular, our lattice model does not do justice to dynamical
charge fluctuations, which should further enhance the role of Coulomb
correlations even in absence of long-range charge order. Indeed, results
supporting the robustness of the intermediate correlated metallic
phase, have been recently obtained in Ref.~{[}\onlinecite{Fratini09}{]}
using complementary methods which emphasized the important role of
longer-range Coulomb interactions. In addition, the influence of disorder
also needs to be addressed in the context of the Wigner-Mott scenario
we propose. These effects can be naturally incorporated in our framework
using the recently developed {}``extended'' DMFT approaches\cite{pankov05prl,RoP2005review}.
This interesting and important task opens an interesting avenue for
future work. 

The authors thank S. Fratini and J. Merino and for useful discussions.
D.T. acknowledges the Serbian Ministry of Science and Technological
Development under Project No.~OI 141035; NATO Science for Peace and
Security Programme Reintegration Grant No. EAP.RIG.983235. V.D. was
supported by the NSF grant DMR-0542026. K.H was supported by the NSF
grant DMR-0746395 and Alfred P. Sloan Fellowship. G.K. was supported
by the NSF grant DMR-0906943.

\bibliography{wigner-mott,vlad10}
 
\end{document}